\pdfoutput=1
\documentclass[]{spie}
\usepackage{graphicx}
\usepackage{caption}
\usepackage{subcaption}
\usepackage{gensymb}
\usepackage{amsmath}
\usepackage{siunitx}
\usepackage{footmisc}
\usepackage{hyperref}

\newcommand{\BigTheta}[1]{\ensuremath{\operatorname{\Theta}\left(#1\right)}}

\DeclareSIUnit{\electron}{e\textnormal{-}}
\DeclareSIUnit{\pixel}{px}

\title{STARS: A software application for the EBEX autonomous daytime star cameras}

\author{Daniel Chapman\supit{a}\thanks{Send correspondences to Daniel Chapman: chappy@phys.columbia.edu, 1-949-632-1806.},
Joy Didier\supit{a}, Shaul Hanany\supit{b}, Seth Hillbrand\supit{a},
Michele Limon\supit{a}, Amber Miller\supit{a}, Britt Reichborn-Kjennerud\supit{a},
Greg Tucker\supit{c}, Yury Vinokurov\supit{d}
\skiplinehalf
\supit{a}Columbia University, New York, NY 10027;\\
\supit{b}University of Minnesota School of Physics and Astronomy, Minneapolis, MN 55455;\\
\supit{c}Brown University, Providence, RI 02912;\\
\supit{d}Carnegie Mellon University, Pittsburgh, Pennsylvania 15213
}
\renewcommand\footnotemark{}

\begin{document}
\maketitle 

\begin{abstract}
The E and B Experiment (EBEX) is a balloon-borne telescope designed to probe polarization signals in the CMB resulting from primordial gravitational waves, gravitational lensing, and Galactic dust emission. EBEX completed an 11 day flight over Antarctica in January 2013 and data analysis is underway. EBEX employs two star cameras to achieve its real-time and post-flight pointing requirements. 
We wrote a software application called STARS to operate, command, and collect data from each of the star cameras, and to interface them with the main flight computer. We paid special attention to make the software robust against potential in-flight failures. 
We report on the implementation, testing, and successful in flight performance of STARS.
\end{abstract}

\keywords{CMB, millimeter-wave telescopes, flight control systems, ballooning, star trackers}

\section{INTRODUCTION}

The E and B Experiment (EBEX) is a stratospheric 
balloon-borne microwave experiment designed to probe polarization signals in the cosmic 
microwave background radiation (CMB).\cite{Oxley04}
It is designed to measure or place an upper limit on the inflationary B-mode signal,
a signal predicted by inflationary theories to be imprinted on the CMB by gravitational waves,\cite{kamionkowski97b}\cite{seljak97}
to detect the effects of gravitational lensing on the polarization of the CMB,\cite{zaldarriaga98}
and to characterize polarized foregrounds.\cite{brandt94}\cite{baccigalupi03}

The EBEX experiment consists of a pointed gondola which supports the telescope, detectors, readout electronics, 
pointing sensors, control motors, control and data acquisition computers, and a telemetry system.
The telescope consists of two mirrors that focus light into a cryogenic receiver, where a rotating half-wave plate 
modulates the signal before it is split by a wire grid polarizer onto two focal planes.
The focal planes contain more than a 1000 transition edge sensor (TES) bolometers that operate at three  
frequency bands centered at 150, 250, and 410 GHz. More detail on the EBEX experiments are
provided by other publications.\cite{ebex_britt}\cite{ebex_kevin}\cite{ebex_milligan}
EBEX completed an 11 day science flight over Antarctica in January 2013.

\subsection{Attitude Control and Pointing Requirements}

The attitude control system (ACS) controls the telescope in real time and collects information for post-flight attitude reconstruction. 
During flight attitude control relies on pointing sensors that provide information to the 'flight control program' (FCP), a program that runs 
on the two redundant flight computers. Using this information FCP constructs a real-time pointing solution and compares it to either 
a pre-defined observing schedule, or to observing commands that are relayed by ground-operators. 
The magnitude of difference between 
an actual and desired attitude generates a control signal to each of two azimuth motors and one elevation motor. 

The real-time pointing requirement of $\sim$\SI{0.5}{\degree} is determined such that it is small compared to the 6{\degree}
diameter of the focal plane.
The post-flight attitude reconstruction requirement of \SI{9}{\arcsecond} is required to make the systematic error from pointing a factor 
of 10 smaller than statistical errors.  

\subsection{Pointing Sensors}

EBEX uses two sets of 3-axis orthogonal fiber-optic rate gyroscopes\footnote{KVH DSP-3000} to measure angular velocity.
The gyros have a measured noise level of  \SI{4}{\arcsecond\per\second\per\hertz^{\frac{1}{2}}} 
and operate at a frequency of \SI{1000}{Hz}. In addition EBEX has 10 absolute pointing sensors as listed in Table \ref{table:list_of_absolute_sensors}\footnote{All
of the sensors listed, except for the elevation encoder, could have been active on at least one more coordinate, though for various reasons these additional
functionalities were not implemented.}.  
The six different types of absolute sensors operate on different principles, and so they have different statistical and systematic uncertainties. 
FCP maintains independent pointing solution streams for each active coordinate of each independent sensor.
Each of these pointing streams is the result of combining the absolute pointing measurements (whenever they are available) with gyroscope measurements
in Kalman filters operating at \SI{100.16}{Hz}, the frequency of the FCP main loop.

    \begin{table}[h]
    \caption{List of EBEX absolute sensors.}
    \label{table:list_of_absolute_sensors}
    \begin{center}       
    \begin{tabular}{lll}
    \hline \rule[-1ex]{0pt}{3.5ex} Sensor & Active Coordinates \\
    \hline \rule[-1ex]{0pt}{3.5ex} star camera 0 & azimuth, elevation
    \\ \rule[-1ex]{0pt}{3.5ex} star camera 1 & azimuth, elevation
    \\ \rule[-1ex]{0pt}{3.5ex} magnetometer 0 & azimuth
    \\ \rule[-1ex]{0pt}{3.5ex} magnetometer 1 & azimuth
    \\ \rule[-1ex]{0pt}{3.5ex} sun sensor 1 & azimuth
    \\ \rule[-1ex]{0pt}{3.5ex} sun sensor 2 & azimuth
    \\ \rule[-1ex]{0pt}{3.5ex} EBEX DGPS & azimuth
    \\ \rule[-1ex]{0pt}{3.5ex} CSBF DGPS & azimuth
    \\ \rule[-1ex]{0pt}{3.5ex} elevation encoder & elevation
    \\ \rule[-1ex]{0pt}{3.5ex} clinometer & elevation
    \\ \hline
    \end{tabular}
    \end{center}
    \end{table}

\subsection{Star Cameras}

The two redundant star cameras are the most precise absolute pointing sensors on EBEX.
They operate by capturing images of the sky and identifying known stars in the field of view. We refer to this process as
finding a pointing solution.
The precision of a star camera solution is primarily determined by the specifications of the physical camera.
Each EBEX star camera contains a \SI{200}{\milli\meter} f/1.8 lens\footnote{Canon EF 200 mm F/1.8 L USM Lens}
and a CCD\footnote{Kodak KAF-1603E image sensor} with 1536x1024 \SI{9}{\micro\meter} pixels.
Using pointing measurements we determined that the star cameras have a precision of \SI{1}{\arcsecond} in 
declination, \SI{1}{\arcsecond} in cross-declination, and \SI{45}{\arcsecond} in roll.
This level of precision is sufficient both to meet the real-time pointing requirement and to meet the post-flight pointing requirement.

Each star camera system consists of a pressurized vessel that contains a digital camera, 
an embedded computer, a hard disk, and various supporting electronics.
The Star Tracking Attitude Reconstruction Software (STARS) is a program that runs on the embedded computers.
STARS downloads images from the camera controller, processes the images to find pointing solutions, and 
serves these solutions to the EBEX flight computers. The design, implementation and operation of STARS are the main 
focus of this paper. 

\section{STARS Design Requirements}

When designing STARS we strove to satisfy a number of requirements as follows.

\begin{itemize}
\item {\bf Frequent in-flight solutions}.  The EBEX star cameras were designed to capture images 
roughly once every $\sim$\SI{40}{\second} due to limitations imposed by the scan strategy and the star camera hardware.
Therefore STARS must solve an appreciable fraction of these images in flight because the error on the star camera 
pointing stream increases with the amount of time that passes from the most recent solution,
and because it is the only way that ground operators can have confidence that the star cameras are collecting 
viable data for post-flight analysis.

\item {\bf Robust operation in daylight conditions}. 
EBEX is designed to fly over Antarctica during the austral summer when the Sun is always above the horizon.
Sky brightness, even at 120,000 ft, adds considerable noise to the images, making the signal-to-noise ratio 
of the sources a primary concern. Other unpredictable features  may also pollute the images, including 
mesospheric clouds, satellites, optical vignetting, and internal reflections. Therefore
STARS must be able to process images with substantial noise and unpredictable features.

\item {\bf Autonomous operation}. 
STARS must be able to operate with limited manual intervention due to the low communication
bandwidth between the ground station and the payload. 
This is complicated by the fact that the software's solving functionality cannot be tested within four months of launch
due to 24-hour daylight conditions in Antarctica,
despite the fact that alterations are made to the flight computer program, which STARS interfaces with,
and that the experiment is disassembled and reassembled for shipping.

\item{\bf Minimal dependencies on other subsystems}.
The star cameras depend on other components of the attitude control system for operation.
They were designed to receive exposure triggering events from the scan control algorithms in FCP,
coarse attitude information from the other pointing sensors, commands from pre-defined observing schedules,
and commands from ground operators using the telemetry system.
However, issues can arise in flight with any number of subsystems\footnote{See
Section \ref{section:in_flight_performance} for examples from the EBEX Antarctic flight.},
so STARS must be robust: it must continue to
collect viable data and provide pointing solutions to FCP in instances where the inputs to the
star camera systems are abnormal.

\end{itemize}

\section{STARS Design Principles and Architecture}
\label{design_principles}

STARS is equipped with a number of features and functionalities that are designed to make it more robust (see Table \ref{table:list_of_notable_features}). 
An increase in features and functionalities often leads to an increase in code complexity that can lead to decrease in 
software reliability. For that reason, special attention was paid to good programming practices. 
We emphasized code readability, used logical abstractions and modularity, and enforced thread-safety.  We also 
implemented extensive testing and performance verifications. As a result of these measures, 
throughout the 11 day flight of EBEX STARS reliably provided pointing solutions and never crashed. 

    \begin{table}[h]
    \caption{List of key functionalities provided by STARS.
    } \label{table:list_of_notable_features}
    \begin{center}       
    \begin{tabular}{lll}
    \hline \rule[-1ex]{0pt}{3.5ex} Notable In-Flight Feature & Discussed In \\ \hline
    \rule[-1ex]{0pt}{3.5ex} robust pattern matching & Section \ref{section:pattern_matching} \\
    \rule[-1ex]{0pt}{3.5ex} robust source finding & Section \ref{section:source_finding} \\
    \rule[-1ex]{0pt}{3.5ex} accurate and abundant video display & Section \ref{section:displaying} \\
    \rule[-1ex]{0pt}{3.5ex} accurate and abundant downlink & Section \ref{section:networking} \\
    \rule[-1ex]{0pt}{3.5ex} selective masking & Section \ref{section:selective_masking} \\
    \rule[-1ex]{0pt}{3.5ex} multiple exposures & Section \ref{section:multiple_exposures} \\
    \rule[-1ex]{0pt}{3.5ex} motion PSF source finding & Section \ref{section:motion_psf} \\
    \rule[-1ex]{0pt}{3.5ex} non-stationary autofocus & Section \ref{section:autofocus} \\
    \hline
    \end{tabular}
    \end{center}
    \end{table}

\subsection{Architecture}

STARS is a cross-platform application written in C++.
During flight it runs on Windows due to lack of driver support in Linux.
During development, testing, and post-flight analysis it is primarily used in Linux.

STARS has multiple threads that can be grouped into four primary categories:

\begin{itemize}
  \item Imaging, which handles lens control, camera control, and image capture either from the camera controller during flight or from file during development, testing, and post-flight analysis.
  \item Solving, which handles the image processing and search algorithm for finding a pointing solution from an image.
  \item Networking, which receives commands from the flight computers and serves back pointing solutions, debugging information, and optionally raw images for redundant storage.
  \item A main thread, which handles display, housekeeping, and shared memory.
\end{itemize}

\subsection{Standard Operation}

\begin{figure}
    \begin{center}
    \begin{tabular}{c}
        \begin{subfigure}{6.75in}
        \includegraphics{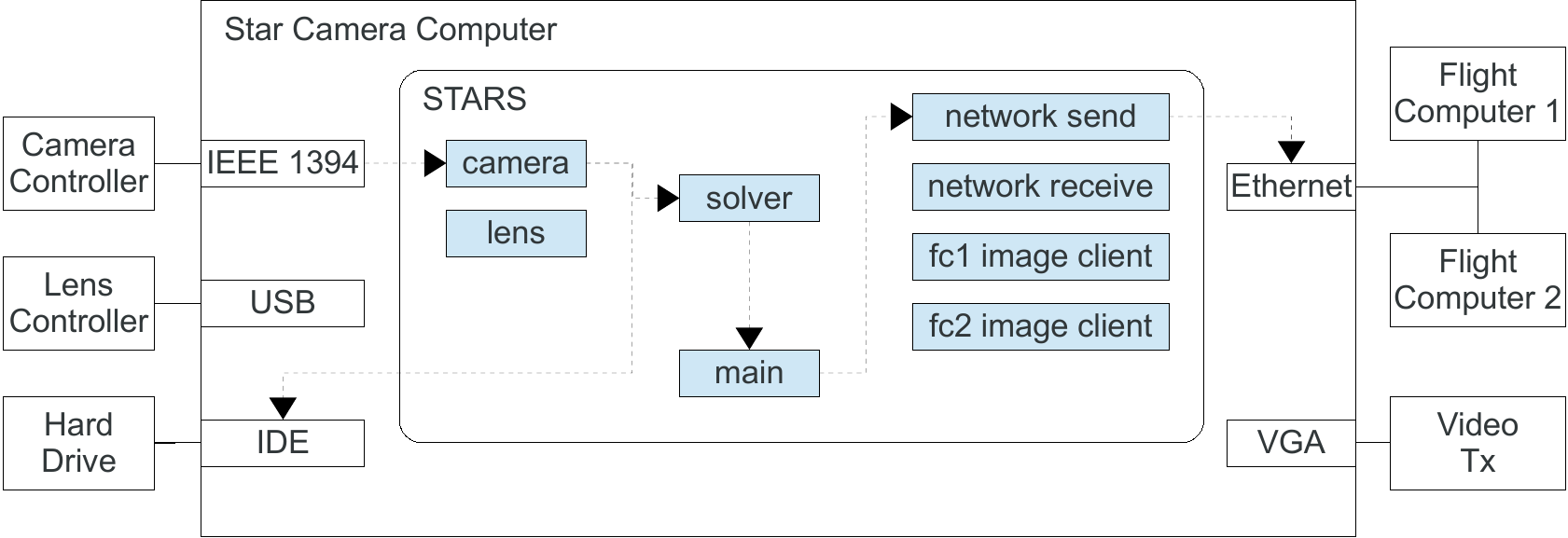}
        \end{subfigure}
    \\ \quad \\
        \begin{subfigure}{6.75in}
        \includegraphics{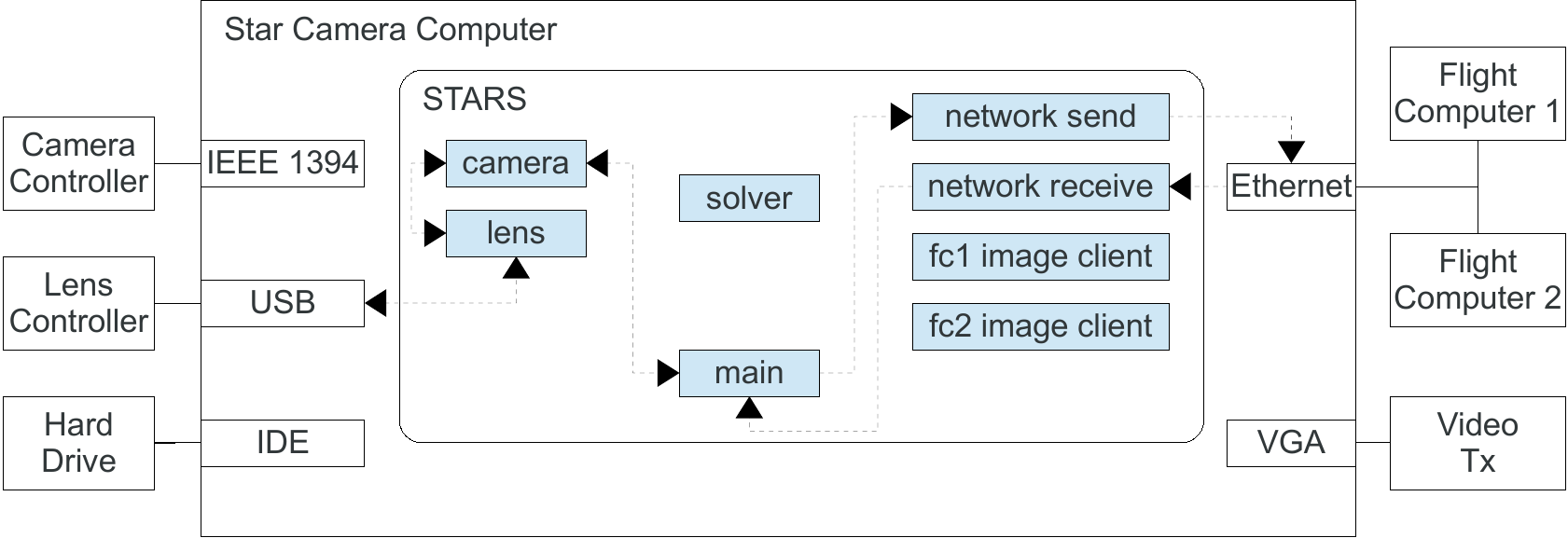}
        \end{subfigure}
    \\ \quad \\
        \begin{subfigure}{6.75in}
        \includegraphics{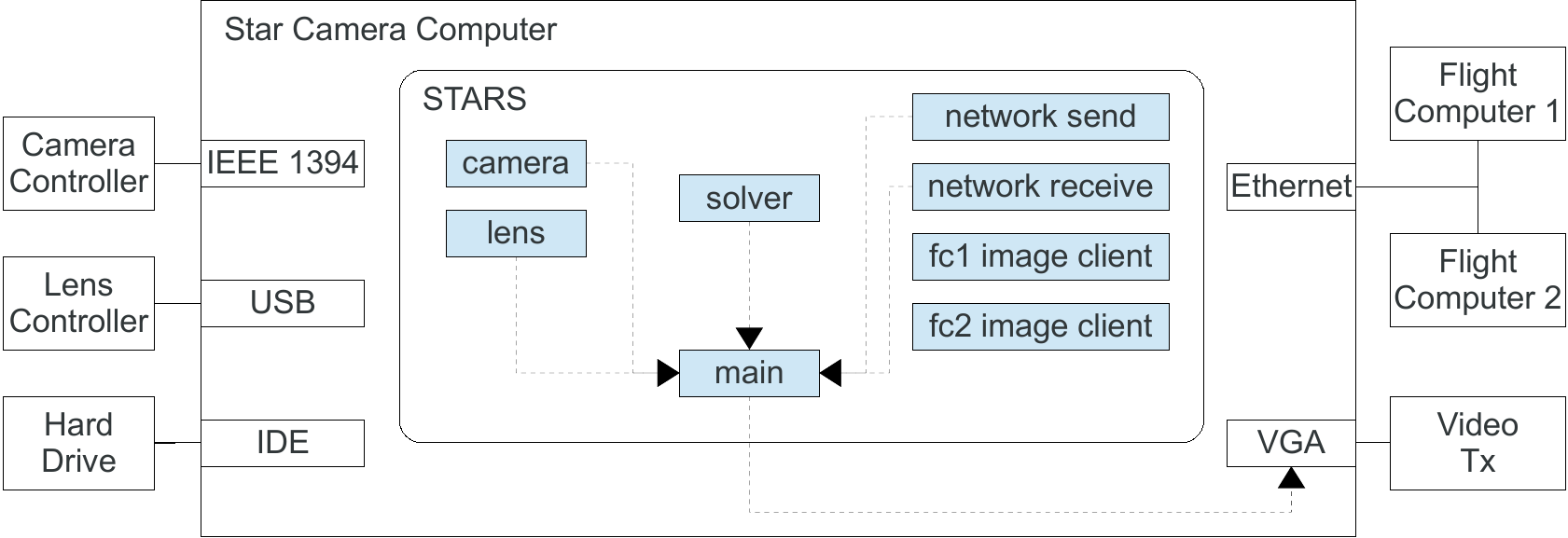}
        \end{subfigure}
    \end{tabular}
    \end{center}
    \caption[leveling]
    {\label{figure:block_diagrams}
    Block diagrams of STARS running on a star camera computer.
    The shaded boxes show the threads that belong to the STARS process.
    Arrows with dotted lines indicate the flow of data.
    Data sent between threads is done safely and efficiently with circular buffers.
    The top panel shows the standard operation: image data and/or pointing solutions make their way from the camera controller to the flight computers.
    The middle panel shows the path that lens requests and results follow from the flight computers to the lens controller and back.
    The bottom panel shows the paths of various objects for display purposes.
    }\end{figure}

As shown in Figure \ref{figure:block_diagrams}, under standard operation an image is downloaded over IEEE 1394 from the camera controller into the camera thread.
The camera thread immediately saves the image to disk and shares the image for the solver thread.  The solver thread receives the image and, if it is ready, processes it to find a pointing solution.  It then shares the pointing solution with the main thread which in turn shares it with the network thread for communication to the flight computers.

\subsection{Shared Memory}

All data shared between threads is done using circular buffers in a thread-safe manner.
Each circular buffer passes data in one direction between exactly two threads.
Since different permutations of threads want to share different kinds of data, there are many possible data objects that can be instantiated via the template circular buffer class for a given path.
This includes raw image objects, lens request objects, pointing solution objects, and many others.
Examples of shared data being passed between threads can be seen in Figure \ref{figure:block_diagrams}.

\subsection{Settings Files}

Settings files allow the user to adjust parameters without recompiling.
Having many configurable settings is useful for development, testing, and post-flight image processing.
However, too many configurable settings can be difficult to keep track of, especially when preparing for launch.
To remedy this, STARS accepts two settings files, one called ``flight.txt'' and one called ``custom.txt''.

The ``flight.txt'' settings file contains all 93 possible configurable parameters, with their default values for flight, and is committed to the repository.
Any parameters in the ``custom.txt'' settings file override those in ``flight.txt''.  The custom file contains only temporary parameters for the purposes of testing and post-flight analysis, along with 8 parameters that must be specified for flight because they are different between the two star cameras\footnote{These parameters are due to hardware differences, e.g. mounted roll, predicted focus position, and platescale.}.
This helps curtail user error by reducing the number of parameters that must be reviewed during the pre-flight checklist from 93 down to 8.

\subsection{Testing}

STARS was developed between 2010 to 2012, and all of its functionality was independently tested as features were implemented.
However, the majority of its testing was due to the fact that many telescope calibration tests,
star camera hardware tests, and tests for other subsystems required fully functional star cameras.
During these tests aberrant software behavior on the part of the star cameras was not tolerated,
in that any unexpected behavior was immediately fixed.

There are two notable features that were implemented to aid with the testing procedure: a sky brightness simulator and a log file parsing utility.

\subsubsection{Brightness Simulator}
\label{feature:brightness_simulator}

The sky brightness simulator is a feature in STARS that injects random Poisson noise into the images to simulate conditions at float.
In the settings file one can specify the photon flux, the camera gain, and the exposure time to inject noise real-time into images
during ground tests.
In flight photon flux predictions have been made using atmospheric transmission modeling software.
This is particularly useful for making fully integrated ground tests as realistic as possible.

\subsubsection{Log Parsing Utility}
\label{feature:log_parsing_utility}

The log parsing utility is a python module that, given a STARS log file path, returns to the user a list of solution python objects.
This expedites development and testing, and is used in post-flight analysis of the images.

\section{STARS Components}

\subsection{Solving - Statistics}

When the solver receives an image, it first calculates statistics on the image
to be used in later steps of image processing,
to be made available to the user,
and to auto-level the image for displaying.
Auto-leveling is necessary because the interesting features in the images only occupy a small fraction of the full image depth.
It is performed by building a histogram of the image and defining the boundaries as the middle \SI{98}{\percent} of the data.
As an optimization, only 1 in every 16 pixels are sampled - evenly spaced on a grid - when building the histogram.

The statistics measured from the image are the mean, noise, gain, and the number of pixels saturated.
The image is broken up into 16x16 pixel cells, and these values are measured for each cell.
The mean of the image is the mean of all the cell means.
The noise of the image is the median of the cell noises; the median is used here to prevent structure within cells (e.g. stars and sharp gradients) or saturated cells from influencing this measurement.
The gain of the image is the mean of the bottom one-sixth of the cell gains for all the ``clean'' cells - cells in which the mean is high enough to be considered significant and in which there are no saturated pixels.

\subsection{Solving - Source Finding}
\label{section:source_finding}

The next step in solving an image is to find the bright spots that may be stars.
Any object identified in an image that could potentially be a star is referred to simply as a ``source'', because it may or may not correspond to an actual star,
while objects loaded from the catalog of known stars are referred to as ``stars'', as they necessarily represent actual stars.
This distinction is important for the next section in which we discuss the pattern matching algorithm,
which attempts to match source objects to star objects.

The source finder primarily operates in two modes: normal and robust, though it also has a third mode called ``motion PSF'' that is discussed in \ref{section:motion_psf}.  Robust mode differs from normal mode in that it can also find sources that are quite out of focus ($\sim$\SI{30}{\pixel} diameter) but at a significant computational cost (\SI{3.8}{\second} on the EBEX star camera computers).

The source finder picks out sources that surpass some tunable threshold above the local level in that region of the image.
To account for the fact that sources span multiple pixels, the source finder picks out sources from a filtered image (smoothed with a Gaussian kernel), and care is taken when calculating the local level in that region of the image.  Thus a pixel is of interest when the following condition is met:

\begin{equation}
\label{equation:peak_condition}
\text{smoothed pixel} > \text{leveled pixel} + \text{threshold}*\text{noise}
\end{equation}

\begin{figure}
    \begin{center}
    \begin{tabular}{c}
    \includegraphics{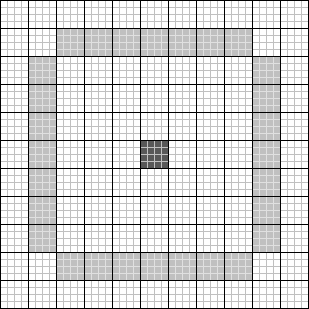}
    \end{tabular}
    \end{center}
    \caption[leveling]
    {\label{figure:leveling}
    The algorithm for determining the regional level around a pixel.
    This image represents a $ 44 \times \SI{44}{px} $ region of an image, where each pixel is shown as a small square with a gray outline.
    The STARS source finder compares a filtered version of the image to a leveled version of the image,
    where the leveled version holds pixel values that represent the regional level of the image.
    To construct the leveled version of the image,
    the 16 pixels in the center, shaded in dark gray, are assigned the mean value of the 448 pixels shaded in light gray.
    The light gray pixels are chosen because they are close enough to the dark gray pixels to be representative of the local level,
    but are far enough away that a potential source centered on a dark gray pixel itself does not bias the level.
    As an optimization, the operation is performed on a coarse version of the image and then upsampled.
    The coarse version of the image is a factor of $ 4 \times 4 $ smaller.
    In this example the coarse pixels are represented by black outlines.
    }\end{figure}

To calculate the local level for a particular pixel,
the source finder takes the mean of a subset of pixels that are between 13 and 24 pixels away from the target pixel.
This is akin to a low-pass filter, but it is performed in position space and excludes pixels that may be biased by the source being searched for.
The specifics of this operation are shown in Figure \ref{figure:leveling},
where each dark gray pixel in the leveled image takes the value of the mean of all the light gray pixels.
This operation is performed coarsely on a $ 4 \times 4 $ downsampled copy of the image as an optimization: 
instead of making $ n \times 449 $ pixel visits, where $n$ is the number of pixels in the full-size image,
it only makes $ n \times 3.8125 $ pixel visits (ignoring boundary cases).

The image is then smoothed with a $ 3 \times \SI{3}{px} $ Gaussian kernel with $ \sigma = \SI{1}{\pixel} $ and the source finder uses this smoothed copy to search for sources.
The smoothing is a 2-D correlation with the Gaussian kernel over the image, and it serves to give weight to sources that extend beyond a single pixel.
The image is broken up into $ 128 \times \SI{128}{px} $ cells\footnote{Many of the numbers used in this section are the default flight values of tunable parameters.\label{footnote:repeat}},
and up to two sources can be selected per cell according to equation \ref{equation:peak_condition}.

If robust mode is enabled, this smoothing and searching procedure is repeated twice more,
once with an $ 11 \times \SI{11}{px} $ kernel with $ \sigma = \SI{3.5}{\pixel} $ and once with a $ 61 \times \SI{61}{px} $ kernel with $ \sigma = \SI{15}{\pixel} $.
As an optimization, each correlation with an $ l \times l $ kernel is broken up into two operations: first a correlation with an $ l \times 1 $ kernel, then a $ 1 \times l $ kernel.
This is mathematically identical if the kernel is Gaussian, and reduces the correlation from \BigTheta{n l^2} to \BigTheta{n l}.

Finally, the sources are sorted by their estimated significance (integrated flux / noise) and the vector of sources is cropped to 10\footref{footnote:repeat}.

Note that we chose to develop robust algorithms for dealing with general artifacts rather than performing a pre-flight flat-fielding procedure.
This is because the image structures seen in flight may not match those seen during controlled tests before flight, which is what actually occurred
(see Section \ref{section:in_flight_performance}).

\subsubsection{Selective Masking Utility}
\label{section:selective_masking}

\begin{figure}
    \begin{center}
    \begin{tabular}{c}
    \includegraphics[width=5in]{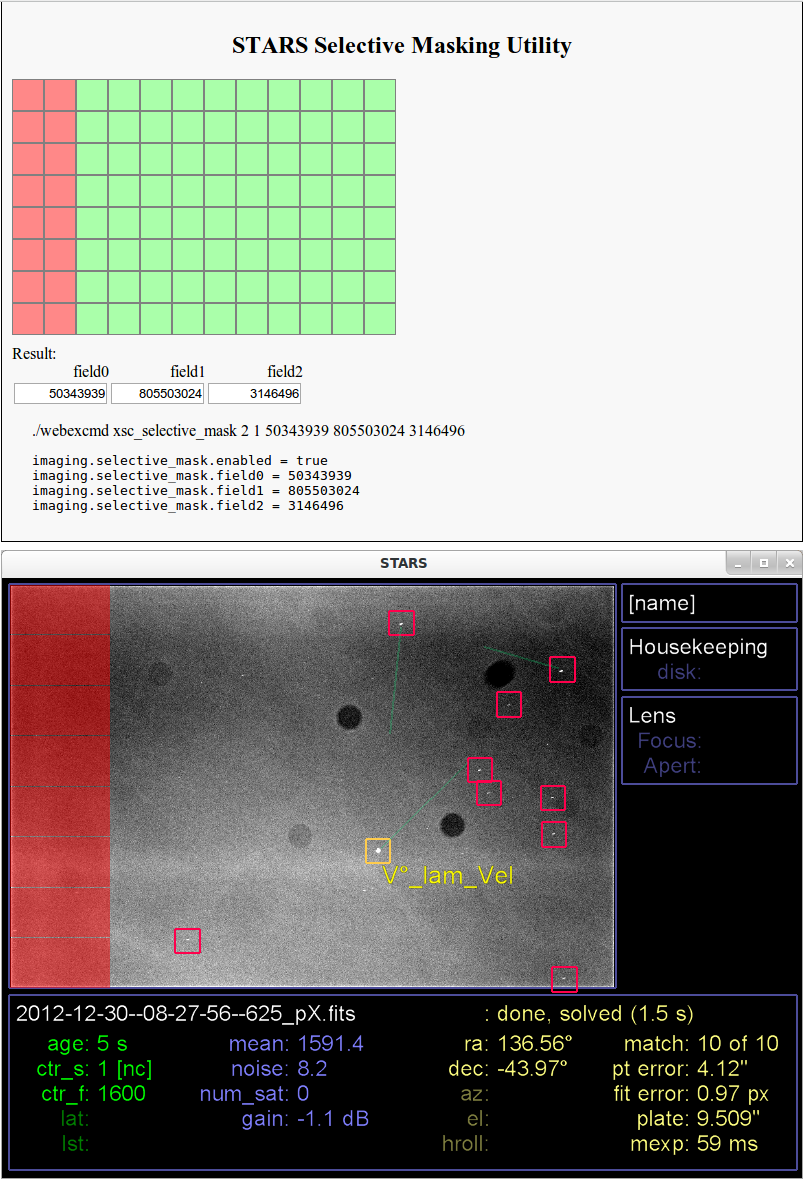}
    \end{tabular}
    \end{center}
    \caption[selective_masking]
    {\label{figure:selective_masking}
    Demonstration of the selective masking utility.
    This html/javascript utility allows users to toggle blocks on or off using the mouse and provides them
    with a command to send to STARS during flight.  STARS then ignores the disabled blocks during source finding.
    In this example the utility is configured to block out the left $\frac{2}{12}$ of the image,
    and the STARS screenshot shows that this area is shaded out of the image, indicating that the source finder
    will not search the left $\frac{2}{12}$ of the image.
    }\end{figure}

To be better prepared for image artifacts that could disrupt the source finder, STARS has the ability to disable $ 128 \times \SI{128}{px} $ blocks of an image.
This is demonstrated in Figure \ref{figure:selective_masking} and was used in flight (see Section \ref{section:in_flight_performance}).

\subsubsection{Motion PSF}
\label{section:motion_psf}

The motion PSF mode is an alternate source finding mode, and is meant to aid in finding sources when the exposure time is too long for the gondola to remain still to within a few pixels.
This is especially common when using multiple exposures.
In this mode, FCP sends to STARS arrays of gyroscope readings for the duration of the exposures.
STARS then reconstructs what the motion blurred source shape should be for each exposure, and uses these shapes as smoothing kernels instead of Gaussian kernels.
Note that individual exposures belonging to a set of multiple exposures must be smoothed separately, but they use kernels that originate from the same starting point so that the streaks for a given star from each exposure contribute to the same source.

\subsection{Solving - Pattern Matching}
\label{section:pattern_matching}

The pattern matcher tries to match the sources in an image to corresponding catalog stars given a particular platescale and set of angles describing the orientation of the image frame using the following procedure:

\begin{enumerate}
\itemsep0em
  \item It picks the brightest three sources in the image.  These three sources form a triangle with known leg lengths (angular distances on the sky).
  \item It selects all the triplets of catalog stars that have similar leg lengths.  For each of these triplets:
    \begin{enumerate}
    \itemsep0em
      \item It performs a least-squares fit to find the pointing solution for the image that would align the selected sources with the selected stars.
      \item Using this pointing solution, it tries to match all the sources in the image to stars from the catalog.  As an optimization it only checks stars in the region of the pointing solution.
      \item With all the possible source-star matches it fits a new pointing solution.
      \item If this tentative pointing solution satisfies all the requirements, it is accepted as the pointing solution.
    \end{enumerate}
  \item If a pointing solution is not accepted, a new triplet or pair of stars is chosen.  This is repeated until every unique triplet combination of the brightest 7 sources is exhausted, and then until every pair combination of two sources is exhausted.
\end{enumerate}

The requirements to decide whether a solution is acceptable may include:

\begin{itemize}
\itemsep0em
  \item a limit on angular distance from a specified horizontal pointing location (azimuth and elevation)
  \item limits on horizontal roll
  \item limits on elevation range, which is convenient for placing constraints based on telescope hardware
  \item a limit on angular distance from a specified equatorial pointing location (right ascension and declination)
  \item limits on equatorial roll
  \item a limit on the pointing solution error
  \item a lower limit on the number of sources matched to stars
\end{itemize}

During flight the limits on pointing coordinates come from estimates of the pointing from the other absolute sensors, which are less accurate.
However, STARS can operate in a lost-in-space mode in which it has no pointing estimate to limit the region of the sky that it searches.
This ``mode'' is activated by disabling all the coordinate-based limits.

The star catalog contains a list of stars (an ID, coordinates, and magnitude for each star) filtered from a list of over one million stars down to approximately 20 per EBEX star camera field of view.
Filtering the catalog to contain approximately the same number of stars per field of view is an important optimization.
If instead the catalog went to a fixed magnitude limit, it would have to go fairly deep to accommodate the sparse regions of sky, which would result in a large number of stars in dense regions near the galactic plane.
Keeping the number of stars in the catalog low reduces the amount of time the search algorithm takes to find a solution.

The catalog is preprocessed into different forms as an optimization for the procedure listed above.
The second step, which involves loading triplets or pairs of stars from the catalog, would be slow if it was to be done dynamically.
Instead the catalog pre-builds every pair and triplet of stars that fit within a star camera field of view.
Although there are more permutations of triplets than pairs, there are statistically fewer catalog triplets that can match a given source triplet than there are catalog pairs to match a given source pair.
As a result, the pattern matcher can search for a given triplet faster than it can search for a pair, but the triplet section of the catalog takes significantly more disk space (and more time to generate during development) than the pair section.
In this manner the STARS pattern matcher is highly influenced by the Pyramid algorithm \cite{pyramid}.
However, the final step in determining whether a solution should be accepted does not involve jumping to a 4-star permutation (pyramid), but instead uses the triangle solution to fit all the remaining sources in the image.

The ability to check all the sources in the image for each triangle allows the user to set a strict requirement on the number of sources matched.
Allowing for an arbitrarily high requirement on the number of matches allows the user to relax other constraints in the images,
such as the platescale limits and the match tolerances that may become too strict under systematic optical distortions.
The STARS pattern matcher is fast enough to run in a lost-in-space mode during flight.

\subsection{Displaying}
\label{section:displaying}

During the first day of flight when the gondola is in range for line-of-sight communication, and during pre-flight testing,
the user can watch the display output of the star camera computer as captured from the VGA port.
This video display can be crucial for debugging any issues with the star camera when the experiment first reaches float altitude.
It is also useful during testing to fully engage the user with the star camera operation.

    \begin{figure}
    \begin{center}
    \begin{tabular}{c}
    \includegraphics[width=6.75in]{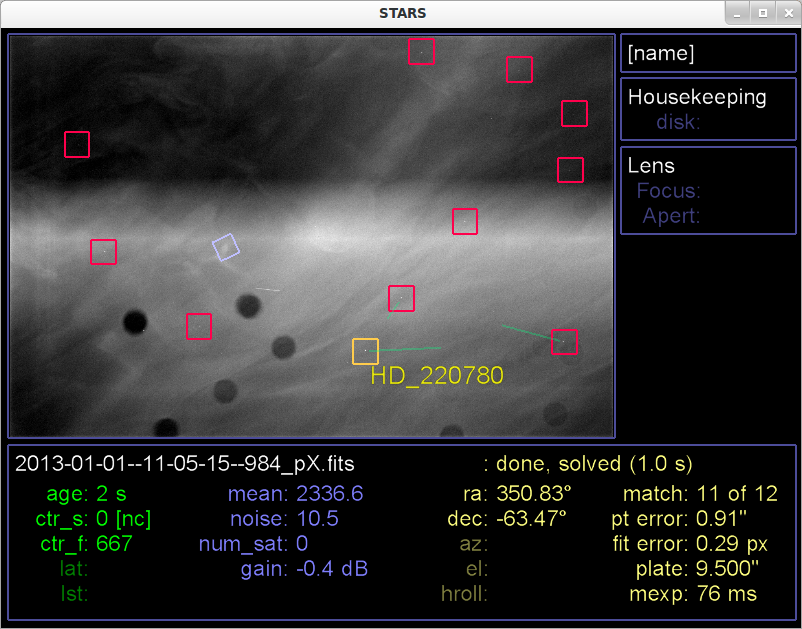}
    \end{tabular}
    \end{center}
    \caption[example\_screenshot]
    {\label{figure:example_screenshot}
    A screenshot of the STARS display, running post-flight on a ground computer on an image from the EBEX Antarctic flight.

    This screenshot shows some of the information displayed to the user during flight.
    On the image itself, the sources are initially identified with rotating (blue) squares.
    When a solution is found, the squares around matched sources change to fixed (red) squares,
    while the brightest of the matched sources is identified by name and a fixed (yellow) square.
    As the pattern matcher searches the catalog for matches from a set of source triplets, the outline of a (green) triangle is traced out as a progress bar connecting the three relevant sources.
    In this case, the solution was found just before making it half way through the potential catalog triplets for the three sources marked by the incomplete (green) triangle.
    If multiple source triplets are tried, old triangles fade away over time in order to not obstruct the image.
    To the right of the image various housekeeping, lens, and network connection information is displayed,
    though not in this post-flight run because it is not being executed on a star camera computer.
    Below is the status information for the image being operated on.
    Administrative information is displayed in the leftmost column, statistics are displayed in the second column from the left,
    and information about the solution found is displayed in the remaining two columns.

    The star camera image shown in this screenshot contains 11 sources matched to catalog stars.
    The image also contains many of the undesirable features seen in flight:
    the dark circles primarily located in the bottom left of the image are dust spots in the optics;
    there is a large gradient due to internal reflections in the optics;
    there is a vertical slab of sharp reflections from the CCD itself on the left side;
    there is a satellite streaking past the image just above one of the dust spots;
    and lastly, this image contains mesospheric clouds.
    The rotating (blue) square represents a source that was not matched; in this case it is a false positive picked out from a piece of cloud.
    }\end{figure}

Figure \ref{figure:example_screenshot} is an example screenshot taken during post-flight processing that shows much of the communication that the display provides.
Not shown in the Figure is additional information that the display can also be commanded to show:
\begin{itemize}
\itemsep0em
  \item A zoomed region of the image.
  \item The status of the autofocus routine, which includes a plot of the focus metrics\footnote{See Section \ref{section:autofocus}} versus the focus position.
  \item The state of all the solution requirements.
\end{itemize}

\subsection{Imaging}

\subsubsection{Camera}

STARS has a camera class that reads images from the camera controller over IEEE 1394, saves them to disk, and then shares them with the solver thread.
A camera instance runs this functionality in its own thread because downloading an image from the camera controller requires a blocking I/O call,
and because the camera controller buffers should be flushed into STARS quickly, before the next set of triggers occurs.

STARS also has a camera class that loads images from disk to be used during testing and post-flight processing.
Only one of these classes is instantiated when STARS runs, as determined by one of the parameters in the settings file.

\subsubsection{Multiple Exposures}
\label{section:multiple_exposures}

STARS has the ability to capture and process multiple exposures taken consecutively.\footnote{The hardware
in the EBEX star cameras allows for up to four exposures separated by \SI{190}{\ms} gaps.}
Multiple exposures are employed when the user would like to capture more light for improved signal-to-noise, but is running up
against the saturation limit of the CCD.
STARS reads multiple exposures and effectively coadds them for image processing.

When coadded multiple exposures are used on EBEX, they tend be to captured over a relatively large amount of time ($\sim$800 to \SI{1300}{\ms})
compared to a normal exposure ($\sim$\SI{300}{\ms}),
so the motion PSF feature discussed in Section \ref{section:motion_psf} is more important in this situation.

\subsubsection{Lens Control}
\label{section:lens_control}

The lens thread controls the focus and aperture by communicating with a physical lens controller over a serial port.
Lens requests typically originate in the networking thread - as ground commands are passed through the flight computers - and terminate in the lens thread.
Once the lens thread has completed the request, the result is sent back to the networking thread to be sent back to the flight computers and ultimately downlinked.

However, the main thread should also know the state of the lens in order to display it to the user,
and the camera thread should also know the state of the lens in order to store it in the saved image headers.
The issue to be overcome is that these threads are asynchronous with the lens thread,
and in fact the lens thread may not even know the state of the lens
between when it sends a command over serial to the physical lens controller and when it receives a confirmation,
which can be up to a couple seconds.
The solution involves counters and passing lens requests/results sequentially.

When the flight computers make a lens request, they also assign it a count.
This request (with a specific count) gets passed into the STARS networking thread, which passes it to the main thread, then the camera thread,
and finally the lens thread which takes action.
Once the action is complete, a result with a corresponding counter gets passed back to the flight computers via the same path.
In this way, for example, if the main thread sees a focus request go to the lens, but has not yet seen the corresponding focus result come back from the lens
(i.e. the counters differ), then it considers the focus value to be unknown, and displays it as such to the user.
This flow of requests and results can be seen in Figure \ref{figure:block_diagrams}.

This is an example of the way memory is shared between processes, and the emphasis placed on sharing accurate information with the user.
The same method is used for the camera gain, which is controlled by the camera thread,
and the status of the autofocus routine, which both originates and terminates in the lens thread
but takes a round-about path through the other threads for this reason.

\subsubsection{Autofocus}
\label{section:autofocus}

The autofocus works by stepping the lens through a range of focus positions and calculating metrics on an image taken at each step.
STARS employs two separate metrics, described below, but will prefer the focus position found by the second metric if it is available.

The first metric is the peak value of the brightest source in the image.
The peak value of a source that represents an actual star is a good metric because, as the focus moves away from the optimal position, the peak value decreases with the square of the width of the source.
If the gondola can point at a fixed patch of sky, such that over the course of an autofocus routine (minutes) the brightest star remains in the frame and another brighter star does not enter the frame, this is an effective metric.

The second metric is the peak value of every source that corresponds to a matched star.
Normally the autofocus is run to find the optimal focus position to maximize the signal-to-noise of the stars, thus enabling the pattern matcher to find a solution.
However, the STARS solver can find solutions relatively far from the optimal focus position.
When a solution is found, the metric for each star identified is stored.
Over the course of many focus steps, STARS actually builds many second metrics, one for each star, and it does so regardless of whether the stars are the brightest objects in the frame or whether the frame moves around.
This metric has the advantage that it does not require a highly stationary gondola to determine the best focus, though it does require the gondola to remain in the same region, or revisit part of the same region, as the focus steps past the optimal point.

\subsection{Networking}
\label{section:networking}

STARS accepts network connections from both flight computers.
Each flight computer attempts to establish a connection to each star camera every 9 seconds if it is not already connected.
Once connected, FCP and STARS send updates to each other once every \SI{0.5}{\second}.
When a gondola operator sends a command from the ground, FCP receives the command and passes it on to STARS over the network connection.
STARS shares solution and other information with both flight computers,
but only accepts commands from the flight computer that is in-charge,
which is a designation assigned by a watchdog timer as the less-recently rebooted flight computer.

Counters are also used to protect STARS from spurious state changes in FCP.
These state changes have been observed many times,
but would require significant changes in FCP to eliminate.
To decrease the probability that STARS accepts a spurious command from FCP that did not originate from a ground command,
it checks a timer to ensure that any particular state change from FCP coincides temporally with the corresponding ground command.
With this safety mechanism, no spurious commands have been observed.

To help with potential in-flight debugging, STARS shares 44 variables with FCP to be downlinked to the ground operator.
These variables include administrative counters, housekeeping measurements, the status of camera and lens parameters,
and statistics and solving information about the current image.

\section{STARS - Successful In-Flight Performance}
\label{section:in_flight_performance}

STARS was used in the EBEX Antarctic flight.
Throughout this 11-day flight STARS never crashed, and it reliably served pointing solutions to the flight computers
with minimal intervention
despite considerable challenges,
which we will discuss in the following paragraphs.
The fact that STARS required minimal intervention was especially important during sparse downlink times on the EBEX flight.

One of the challenges that STARS handled involved misinformation from other subsystems.
A system clock failure on one of the flight computers and eventual failures of all three GPS systems
caused FCP to share incorrect pointing information with STARS for various reasons and at various times throughout the flight.
Nevertheless, STARS continued to provide pointing solutions due to the
fast lost-in-space pattern matching capabilities discussed in Section \ref{section:pattern_matching}.

Due to an issue with an azimuth motor the gondola was unable to remain stationary for more than one second.
This is necessary to perform safe re-focusing procedures, in which we manually change the focus by small amounts
to find the best focus position.\footnote{This
is preferred over the autofocus routine which incurs more risk because it changes the mechanical focus by larger amounts.
Considering this risk, and that the source finder continued to work on the out-of-focus sources,
we opted not to test the non-stationary autofocus feature in flight.}
Thus, as a result of thermal variations in the optics,
the sources in the images expanded and contracted from 2 px to 12 px in diameter over the course of hours and days.
Nevertheless, STARS continued to solve on these out-of-focus sources
due to the source finder's smoothing and leveling procedure discussed in Section \ref{section:source_finding},
and because the resulting loss in signal-to-noise was recovered
by the multiple exposures feature discussed in Section\ref{section:multiple_exposures}.

Another consequence of the azimuth motor issue was that the acceleration at scan end-points and the amount of time in between end-points
was different than in the original scan design.
STARS continued to operate in these new conditions due to the architecture described in Section \ref{design_principles}
in which the camera and solver threads operate asynchronously on data whenever it becomes available,
independent of the gondola's scan state.

Finally, STARS prevailed against various image artifacts:
dust spots,
mesospheric clouds,
satellites,
optical vignetting,
external reflections from the Sun,
sharp internal reflections of the CCD,
and broad gradients from internal reflections.
The leveling and flux sorting procedures discussed in Section \ref{section:source_finding},
along with the selective masking feature (Section \ref{section:selective_masking}),
helped limit the number of false positives extracted by the source finder.
The false positive sources that did get through did not prevent the pattern matcher from finding a solution due to the
fast search algorithm, which allowed for many triplets to be tested in a matter of seconds,
and due to the stringent criteria on the number of sources matched to stars (Section \ref{section:pattern_matching}).
The multiple exposures feature enabled the source finder to find enough sources to meet this stringent matching criteria
due to the increase in signal-to-noise.

\acknowledgments

EBEX is a NASA supported mission through grant numbers NNX08AG40G and NNX07AP36H. We thank Columbia Scientific Balloon Facility for their enthusiastic support of EBEX. We also acknowledge support from NSF, CNRS, Minnesota Super Computing Institute, Minnesota and Rhode Island Space Grant Consortia, the Science and Technology Facilities Council in the UK, Sigma Xi, and funding from collaborating institutions. This research used resources of the National Energy Research Scientiﬁc Computing Center, which is supported by the office of Science of the U.S. Department of Energy under contract No. DE-AC02-05CH11231.
The McGill authors acknowledge funding from the Canadian Space Agency, Natural Sciences and Engineering Research Council, Canadian Institute for Advanced Research, Canadian Foundation for Innovation and Canada Research Chairs program.

\bibliography{report}   
\bibliographystyle{spiebib}   

\end{document}